\documentclass[fleqn,floatfix,showkeys]{revtex4}

\usepackage{amsmath}
\usepackage{graphicx}

\begin{document}

\title{Ro-vibrational states of H$_2^+$. Variational calculations.}

\author{Vladimir I. \surname{Korobov}}
\affiliation{Bogoliubov Laboratory of Theoretical Physics, Joint Institute
for Nuclear Research, Dubna 141980, Russia}
\email[Contact V.I.~Korobov: ]{korobov@theor.jinr.ru}

\begin{abstract}
The nonrelativistic variational calculation of a complete set of ro-vibrational states in the H$_2^+$ molecular ion supported by the ground $1s\sigma$ adiabatic potential is presented. It includes both bound states and resonances located above the $n\!=\!1$ threshold. In the latter case we also evaluate a predissociation width of a state wherever it is significant. Relativistic and radiative corrections are discussed and effective adiabatic potentials of these corrections are included as supplementary files.
\end{abstract}

\keywords{hydrogen molecular ion, variational methods, bound and quasi-bound states}

\maketitle

\section{Introduction}

For many years it was thought that reliable calculations of energy levels of bound and quasi-bound states in the hydrogen molecular ion may be performed only within the adiabatic approximation with some nonadiabatic corrections \cite{BO,BOmc,Wolnewicz,Carrington89,Moss93}. In the latter work \cite{Moss93} Moss calculated 462 ro-vibrational states of 481 states supported by the ground electronic potential curve using adiabatic approximation with a transformed Hamiltonian and an artificial-channels scattering method.

In the past two decades methods to compute the energy-level structure of H$_2^+$ and HD$^+$ ions, which do not rely on the Born-Oppenheimer approximation, have been intensively developed \cite{Yan99,var99,var00,Hilico00,Frolov1995,Frolov02}. These are diverse \emph{ab initio} approximations based on variational expansions of the nonrelativistic three-body wave function. Eventually, it has been shown that the ground state of H$_2^+$ ion may be calculated with as much as 34 significant digits \cite{Yan07,Yan14}.

On the other hand, in experiment, the two quasi-bound states have been observed recently \cite{Merkt16}, which were not accessible in the Moss calculations \cite{Moss93}. Along with these two states energies of a series of weakly bound and some low vibrational states have been measured by using the pulsed-field-ionization zero-kinetic-energy photoelectron spectroscopy \cite{Merkt16JMS}. This study paves a road for spectroscopy of a wide range of states in the H$_2^+$ ion, where the number of successful experiments is drastically smaller than for the HD$^+$ ion due to absence of the electric dipole moment in H$_2^+$ allowing rotational-vibrational transitions.

By our work we want to demonstrate that all the 481 levels that are supported by the ground electronic potential curve may be calculated variationally. More over we claim that for all the bound states we are able to receive the nonrelativistic binding energy with precision of at least $10^{-7}$ cm$^{-1}$ and that for resonant states lying above the dissociation threshold a width ($\Gamma$) as well as the position energy ($E_r$) is obtained with high precision. Thus by using the \emph{ab initio} variational method we have managed to cover the whole realm of existing states related to the $1s\sigma$ electronic adiabatic potential.

In what follows in calculations we adopt the CODATA14 \cite{CODATA14} values for physical constants. Atomic units are used throughout.

\section{Method}

In our studies, the stationary states in the H$_2^+$ molecular ion are determined by the non-relativistic Schr\"odinger equation for three particles:
\begin{equation}\label{eq:NR}
\begin{array}{@{}l}\displaystyle
(H_{0}-E_0)\Psi_{0}=0,
\\[3mm]\displaystyle
H_{0}=
   -\frac{1}{2M}\mbox{\ensuremath{\nabla}}_{1}^{2}
   -\frac{1}{2M}\mbox{\ensuremath{\nabla}}_{2}^{2}
   -\frac{1}{2m_{e}}\mbox{\ensuremath{\nabla}}^{2}
   +\frac{1}{R}-\frac{1}{r_{1}}-\frac{1}{r_{2}}.
\end{array}
\end{equation}
Here $M$ is a proton mass, $R$ is the internuclear distance, $r_{1}$ and $r_{2}$ are the distances from nuclei 1 and 2 to the electron, respectively. The state $\Psi_{0}=|v\, N\rangle$ is characterized by the vibrational and rotational quantum numbers $v,\, N$, and $E_{0}$ is its energy.

\subsection{Variational exponential expansion}

Wave functions of rotational-vibrational states in the molecular hydrogen positive ion are approximated by the variational exponential expansion, which has been successfully exploited and developed by many authors and, in particular, in Ref.~\cite{var99,var00}. More precisely, the wave function for a state with a total orbital angular momentum $N$ and of a total spatial parity $\pi=(-1)^{N}$ is expanded as follows:
\begin{equation}\label{eq:exp}
\begin{array}{@{}l}\displaystyle
\Psi_{NM}^{\pi}(\mathbf{R},\mathbf{r}_{1}) =
   \sum_{l_{1}+l_{2}=N}
      \mathcal{Y}_{NM}^{l_{1}l_{2}}(\mathbf{R},\mathbf{r}_{1})
      G_{l_{1}l_{2}}^{N\pi}(R,r_{1},r_{2}),
\\[3mm]\displaystyle\hspace{25mm}
G_{l_{1}l_{2}}^{N\pi}(R,r_{1},r_{2}) =
   \sum_{n=1}^{N_{\rm max}}
      \Big\{
         C_{n}\,\mbox{Re}\bigl[e^{-\alpha_{n}R-\beta_{n}r_{1}-\gamma_{n}r_{2}}\bigr]+
         D_{n}\,\mbox{Im}\bigl[e^{-\alpha_{n}R-\beta_{n}r_{1}-\gamma_{n}r_{2}}\bigr]\Big\}\,,
\end{array}
\end{equation}
where ${\cal{Y}}^{l_1,l_2}_{NM}(\mathbf{r}_1,\mathbf{r}_2)$ are the solid bipolar harmonics defined as in Ref.~\cite{Var88},
\[
\mathcal{Y}^{l_1,l_2}_{NM}(\mathbf{r}_1,\mathbf{r}_2) =
   r_1^{l_1}r_2^{l_2}\left\{Y_{l_1}\otimes Y_{l_2}\right\}_{NM},
\]
and $N$ is the total orbital angular momentum of a state. Complex parameters $\alpha_k$, $\beta_k$, and $\gamma_k$ are generated in a quasirandom way \cite{Frolov1995,var99}:
\begin{equation}
\begin{array}{@{}l}\displaystyle
\alpha_k =
   \left[\left\lfloor\frac{1}{2}k(k+1)\sqrt{p_{\alpha}}\right\rfloor(A_2-A_1)+A_1\right]
   +i\left[\left\lfloor\frac{1}{2}k(k+1)\sqrt{q_{\alpha}}\right\rfloor(A'_2-A'_1)+A'_1\right]\,,
\end{array}
\end{equation}
where $\lfloor{x}\rfloor$ designates the fractional part of $x$, $p_{\alpha}$ and $q_{\alpha}$ are some prime numbers, and $[A_1,A_2]$ and $[A'_1,A'_2]$ are real variational intervals, which need to be optimized. Parameters $\beta_k$ and $\gamma_k$ are obtained in a similar way. The use of complex exponents instead of real ones is dictated by the oscillatory behavior of the vibrational part of the wave function and, as it has been established empirically, essentially improves the convergence rate for the energy of a state. Other details of the method, such as the use of a multilayer structure to optimize the trial wave function, may be found in \cite{var00}.

Some words should be added related to a choice of the coordinate system. The two position vectors $\mathbf{R}$ and $\mathbf{r}_1$ are taken as basis vectors for the angular part of expansion (\ref{eq:exp}). That does not allow to use the apparent symmetry of permutation of two protons as identical particles or makes it too difficult to realize. On the other hand, summation over the angular part of the wave function in Eq.~(\ref{eq:exp}) converges very rapidly to an exact wave function with increase of $l_2$, since this summation has close connection with the sum over the azimuthal quantum number $m$, see \cite{var99}. Generally it is enough to keep three components with $l_2=0,1,2$ in the expansion to provide the energy as accurate as 16 significant digits. For a large total orbital angular momentum, $N$, it gives very serious gain in computation time than the ability of explicit symmetrization of the wave function.

\begin{table}[t]
	\begin{center}
		\caption{Nonrelativistic energies ($E = E_r + i\Gamma/2$) for the resonant states supported by the adiabatic $1s\sigma$ curve in the $\mbox{H}_2^+$ molecular ion. The energies are computed relative to the 1$S$ hydrogen atom threshold. The states in the first three lines are label by $(v,N)$.} \label{H2+_res}
		\hspace*{-5mm}\footnotesize
		\begin{tabular}{r@{\hspace{3mm}}c@{\hspace{3mm}}c@{\hspace{3mm}}c@{\hspace{3mm}}c@{\hspace{3mm}}c}
			\hline\hline
			& (18,4) & (17,7) & (16,10) & (15,12) & (15,13)  \\
			& $1.84327(4)\!+\!i\,0.09561$      & $11.0572(1)\!+\!i\,0.0810$ & $42.216(3)\!+\!i\,0.456$ &
			$40.90159(2)\!+\!i\,0.00081$ & $112.480(1)\!+\!i\,4.247$ \\
			\hline
			& (14,14) & (14,15) & (13,16) & (13,17) & (12,18)  \\
			& $41.631840(1)\!+\!i\,0.000001$ & $148.7111(3)\!+\!i\,1.2509$ & $56.921993$ &
			$197.71413(2)\!+\!i\,0.47050$ & $95.158759$ \\
			\hline
			& (12,19) & (11,20)   \\
			& $266.92302(1)\!+\!i\,0.32824$ & $162.051960$ \\
			\hline\hline
			$v$ & $N=21$ & $N=22$ & $N=23$ & $N=24$ & $N=25$ \\
			8 &   &   &   &   & 255.720129 \\
			9 &   &   & 109.433662 & $395.57351\!+\!i\,0.00029$ & $640.3459(1)\!+\!i\,1.8701$ \\
			10 & 2.239830 & $261.388316\!+\!i\,000018$ & $485.57281\!+\!i\,0.80492$ \\
			11 & $361.74636\!+\!i\,0.41981$ \\
			\hline
			$v$ & $N=26$ & $N=27$ & $N=28$ & $N=29$ & $N=30$ \\
			5 &   &   &   & 85.558635 & 526.350160 \\
			6 &   &   & 279.938459 & $668.75053\!+\!i\,0.00001$ & $1018.2881(3)\!+\!i\,0.2275$ \\
			7 & 76.533945 & 441.998395 & $773.51485\!+\!i\,0.03352$ & $1047.1282\!+\!i\,9.9532$ \\
			8 & $566.03499\!+\!i\,0.00361$ & $827.1839(1)\!+\!i\,4.4709$ \\
			\hline
			$v$ & $N=31$ & $N=32$ & $N=33$ & $N=34$ & $N=35$ \\
			1 &   &   &   &   & 283.727120 \\
			2 &   &   &   & 485.913428 & 1021.058997 \\
			3 &   & 151.261628 & 664.321137 & $1148.75419\!+\!i\,0.00001$ & $1595.6407(1)\!+\!i\,0.0529$ \\
			4 & 352.517940 & 815.863900 & $1245.00401\!+\!i\,0.00627$ & $1620.9184(2)\!+\!i\,3.7466$ \\
			5 & $936.28999\!+\!i\,0.00040$ & $1300.4846(3)\!+\!i\,1.1007$ \\
			\hline
			$v$ & $N=36$ & $N=37$ & $N=38$ & $N=39$ & $N=40$ \\
			0 &  60.201298 & 688.802131 & 1293.849345 & 1872.161280 & $2418.98845\!+\!i\,0.00341$ \\
			1 & 866.659020 & 1423.264207 & $1948.25079\!+\!i\,0.00109$ & $2430.70205\!+\!i\,0.72057$ & $2849.7390\!+\!i\,23.3570$ \\
			2 & $1525.72814\!+\!i\,0.00012$  & $1989.77978\!+\!i\,0.25777$ & $2389.4651\!+\!i\,16.8142$ \\
			3 & $1982.2541\!+\!i\,9.1690$ \\
			\hline
			$v$ & $N=41$ \\
			0 & $2924.53309\!+\!i\,0.99382$ \\
			\hline\hline
		\end{tabular}
	\end{center}
\end{table}

\subsection{Resonances and the Complex Coordinate Rotation method}

Beyond bound states in H$_2^+$ we also consider states, which are above the ground state dissociation threshold and thus may dissociate by penetrating through the centrifugal potential barrier: $V_{rot}=\frac{N(N+1)}{MR^2}$. For this case we have to use some formalism, which may rigorously treats resonances. Such a tool for variational methods, that is most efficient and simple in a practical use, is the Complex Coordinate Rotation (CCR) method. We briefly describe it below.

The Coulomb Hamiltonian (\ref{eq:NR}) is an analytic function of the dilatation parameter $\theta$
\begin{equation}\label{dilatation}
\left(U(\theta)f\right)(\mathbf{r}) = e^{d\theta/2}f(e^\theta\mathbf{r}),
\qquad H(\theta)=U(\theta)HU^{-1}(\theta),
\end{equation}
for real $\theta$, or in other words may be expanded in a convergent power series of the dilatation parameter $\theta$ on some open interval, and thus can be analytically continued to the complex plane. Parameter $d$ in Eq.~(\ref{dilatation}) is a dimension of the coordinate space, say, for a single electron in a three-dimensional space: $d=3$.

The Complex--Coordinate Rotation method \cite{Reinhardt82,Ho83} "rotates" the coordinates of the dynamical system ($\theta=i\varphi$), $r_{ij}\rightarrow r_{ij} e^{i\varphi}$, where $\varphi$ is the parameter of the complex rotation. Under this transformation the Hamiltonian
(\ref{eq:NR}) changes as a function of $\varphi$
\begin{equation}\label{rotHam}
H_{\varphi} = T e^{-2 i \varphi} + V e^{-i \varphi},
\end{equation}
where $T$ and $V$ are the kinetic energy and Coulomb potential operators. The continuum spectrum of $H_{\varphi}$ is rotated on the complex plane around branch points ("thresholds") to "uncover" resonant poles situated on the unphysical sheet of the Reimann surface in accordance with the Augilar-Balslev-Combes theorem \cite{ABC}. The resonance energy is then determined by solving the complex eigenvalue problem for the "rotated" Hamiltonian
\begin{equation}
(H_{\varphi} - E)\Psi_{\varphi} = 0, \label{roteqn}
\end{equation}
The eigenfunction $\Psi_{\varphi}$ obtained from Eq.~(\ref{roteqn}), is square-integrable and the corresponding complex eigenvalue $E = E_r - i\Gamma/2$ defines the energy $E_r$ and the width of the resonance, $\Gamma$, the latter is being related to the Auger rate as $\lambda_A = \Gamma/\hbar$.

The use of a finite set of $N$ basis functions defined by (\ref{eq:exp}) reduces the problem (\ref{roteqn}) to the generalized algebraic complex eigenvalue problem
\begin{equation}
(A-\lambda B) x = 0, \label{gaevalp}
\end{equation}
where $A=\langle\Psi_{\varphi}|H_{\varphi}|\Psi_{\varphi}\rangle$ is the finite $N\times N$ matrix of the Hamiltonian in this basis, and $B$ is the matrix of overlap
$B=\langle\Psi_{\varphi}|\Psi_{\varphi}\rangle$, and $x$ is a vector of linear coefficients $C_n$ and $D_n$ from Eq.~(\ref{eq:exp}).

\begin{figure}[t]
\begin{center}
\includegraphics[width=0.55\textwidth]{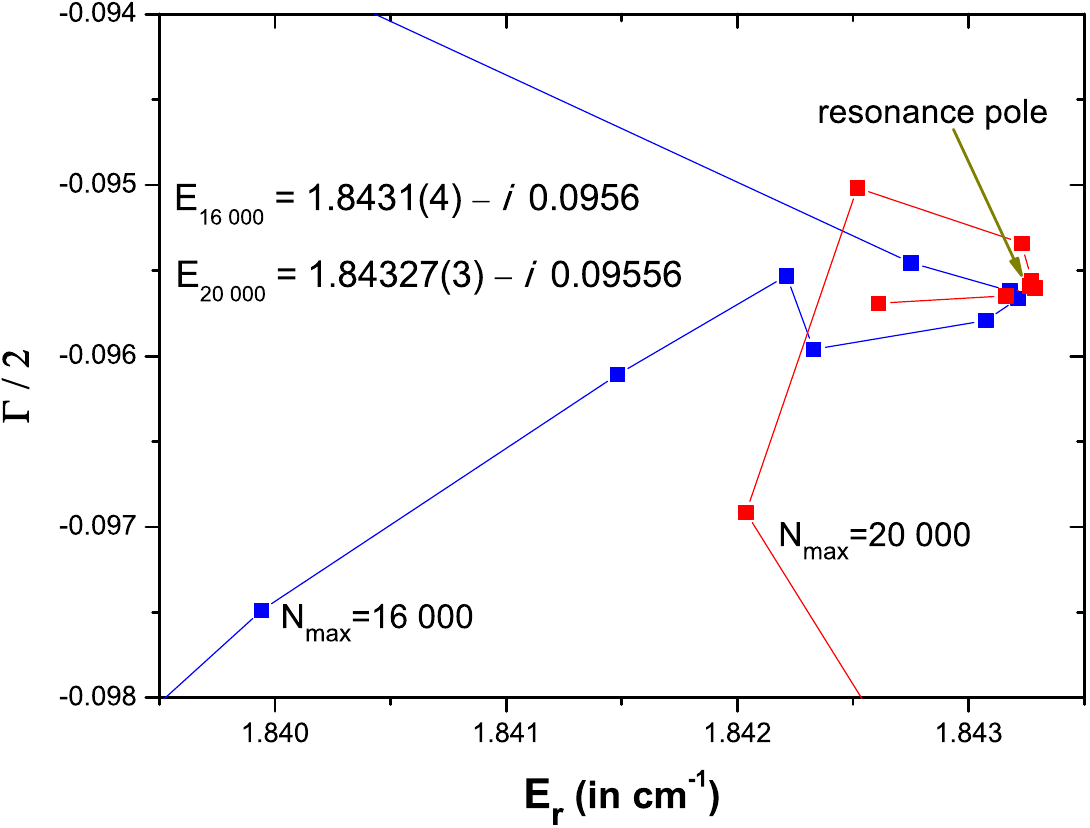}
\end{center}
\caption{Rotational paths and position of the resonance pole, $E_r-i\Gamma/2$ for the first rotational quasi-bound state: $(v=18,N=4)$.}\label{fig1}
\end{figure}

An example of practical calculation is given in Fig.~\ref{fig1}. The two rotational paths for a basis set of $N_{\rm max}=16\,000$ and $N_{\rm max}=20\,000$ are shown on the plot. A step in the rotational angle $\varphi$ between two sequential calculations is constant and equal to $\Delta\varphi=0.02$. A point where the paths become stabilized determines a position of the resonance pole.

\begin{table}[t]
\begin{center}
\caption{Nonrelativistic dissociation energies (in cm$^{-1}$) for the bound states supported by the adiabatic $1s\sigma$ curve in the $\mbox{H}_2^+$ molecular ion.} \label{H2+}
\hspace*{-1mm}\footnotesize
\begin{tabular}{r@{\hspace{3mm}}r@{\hspace{3mm}}r@{\hspace{3mm}}r@{\hspace{3mm}}r@{\hspace{3mm}}r@{\hspace{3mm}}r@{\hspace{3mm}}r@{\hspace{3mm}}r}
\hline\hline
 $v$ & $N=0$~~~~~ & $N=1$~~~~~ & $N=2$~~~~~ & $N=3$~~~~~ & $N=4$~~~~~ & $N=5$~~~~~ & $N=6$~~~~~ & $N=7$~~~~~ \\
\hline
 0 & 21379.2923402 & 21321.0603885 & 21205.0607384 & 21032.2098693 & 20803.8531565
   & 20521.7316112 & 20187.9405636 & 19804.8824805 \\
 1 & 19188.1928214 & 19133.0260782 & 19023.1374588 & 18859.4050711 & 18643.1177275
   & 18375.9428559 & 18059.8867005 & 17697.2489649 \\
 2 & 17124.3028404 & 17072.0963314 & 16968.1099327 & 16813.1856561 & 16608.5591595
   & 16355.8288158 & 16056.9174116 & 15714.0285801 \\
 3 & 15183.3996383 & 15134.0638326 & 15035.8016848 & 14889.4212934 & 14696.1084324
   & 14457.3967712 & 14175.1310533 & 13851.4252936 \\
 4 & 13361.9186852 & 13315.3792105 & 13222.6936295 & 13084.6382503 & 12902.3521462
   & 12677.3085241 & 12411.2793744 & 12106.2954148 \\
 5 & 11656.9363868 & 11613.1340029 & 11525.9075567 & 11396.0036645 & 11224.5178524
   & 11012.8670746 & 10762.7558352 & 10476.1378741 \\
 6 & 10066.1625524 & 10025.0534095 &  9943.1994567 &  9821.3197386 &  9660.4694260
   &  9462.0134951 &  9227.5943335 &  8959.0951828 \\
 7 &  8587.9429632 &  8549.4991884 &  8472.9630777 &  8359.0283032 &  8208.7129989
   &  8023.3346260 &  7804.4790931 &  7553.9659925 \\
 8 &  7221.2727634 &  7185.4834394 &  7114.2445059 &  7008.2266154 &  6868.4144298
   &  6696.0827128 &  6492.7670221 &  6260.2308134 \\
 9 &  5965.8218314 &  5932.6944717 &  5866.7690380 &  5768.6957768 &  5639.4298549
   &  5480.2087486 &  5292.5246048 &  5078.0933447 \\
10 &  4821.9738235 &  4791.5364977 &  4730.9822003 &  4640.9438021 &  4522.3516105
   &  4376.4121705 &  4204.5824579 &  4008.5412090 \\
11 &  3790.8811854 &  3763.1855389 &  3708.1074836 &  3626.2662009 &  3518.5728178
   &  3386.2108081 &  3230.6122959 &  3053.4320053 \\
12 &  2874.5390152 &  2849.6645766 &  2800.2241396 &  2726.8279323 &  2630.3752543
   &  2512.0368083 &  2373.2336051 &  2215.6142458 \\
13 &  2075.8807566 &  2053.9411944 &  2010.3688604 &  1945.7714956 &  1861.0467218
   &  1757.3669066 &  1636.1615989 &  1499.0995466 \\
14 &  1398.8964633 &  1380.0489646 &  1342.6636898 &  1287.3553880 &  1215.0351262
   &  1126.8989112 &  1024.4156178 &   909.3168690 \\
15 &   848.7622364 &   833.2221873 &   802.4620268 &   757.1188031 &   698.1419406
   &   626.7885089 &   544.6216071 &   453.5167624 \\
16 &   431.9115198 &   419.9765224 &   396.4497403 &   362.0185899 &   317.7184980
   &   264.9437664 &   205.4739575 &   141.5337864 \\
17 &   155.6515739 &   147.7382066 &   132.3058791 &   110.1557416 &    82.5281627
   &    51.1795292 &    18.5875053 &  \\
18 &    24.0527340 &    20.6202178 &    14.2559364 &    6.0384695 &  &  &  &  \\
19 &     0.7442247 &     0.2210596 &  &  &  &  &  &  \\
\hline
 $v$ & $N=8$~~~~~ & $N=9$~~~~~ & $N=10$~~~~~ & $N=11$~~~~~ & $N=12$~~~~~ & $N=13$~~~~~ & $N=14$~~~~~ & $N=15$~~~~~ \\
\hline
 0 & 19375.2162271 & 18901.8050136 & 18387.6650482 & 17835.9165950 & 17249.7387515
   & 16632.3288692 & 15986.8671684 & 15316.4867788 \\
 1 & 17290.5741515 & 16842.6017735 & 16356.2173874 & 15834.4060704 & 15280.2095848
   & 14696.6880901 & 14696.6880901 & 13453.8085120 \\
 2 & 15329.6001688 & 14906.2566510 & 14446.7624611 & 13953.9778039 & 13430.8181173
   & 12880.2179922 & 12305.1000080 & 11708.3486520 \\
 3 & 13488.6181349 & 13089.2274092 & 12655.9057152 & 12191.3984974 & 11698.5057452
   & 11180.0480687 & 10638.8375751 & 10077.6536958 \\
 4 & 11764.6034106 & 11388.6228529 & 10980.9037425 & 10544.0869022 & 10080.8678857
   &  9593.9651992 &  9086.0932361 &  8559.9400693 \\
 5 & 10155.1754436 &  9802.1980928 &  9419.6626500 &  9010.1157693 &  8576.1600681
   &  8120.4245412 &  7645.5396447 &  7154.1172030 \\
 6 &  8658.6013846 &  8328.3612937 &  7970.7484889 &  7588.2266071 &  7183.3177920
   &  6758.5754345 &  6316.5616060 &  5859.8293798 \\
 7 &  7273.8118656 &  6966.1933074 &  6633.4114973 &  6277.8594460 &  5901.9929424
   &  5508.3058859 &  5099.3104566 &  4677.5224047 \\
 8 &  6000.4308226 &  5715.4824946 &  5407.6270109 &  5079.2012012 &  4732.6113382
   &  4370.3115616 &  3994.7874977 &  3608.5455488 \\
 9 &  4838.8223407 &  4576.7784065 &  4294.1576535 &  3993.2585243 &  3676.4590827
   &  3346.1994587 &  3004.9702655 &  2655.3078880 \\
10 &  3790.1592015 &  3551.4702360 &  3294.6444163 &  3021.9651541 &  2735.8111787
   &  2438.6448030 &  2133.0078588 &  1821.5272023 \\
11 &  2856.5206975 &  2641.8999254 &  2411.7398638 &  2168.3419311 &  1914.1279914
   &  1651.6382626 &  1383.5408828 &  1112.6578630 \\
12 &  2041.0325529 &  1851.5276202 &  1649.3084743 &  1436.7458328 &  1216.3741160
   &   990.9083640 &   763.2839428 &   536.7341786 \\
13 &  1348.0726495 &  1185.1846029 &  1012.7476712 &   833.2923990 &   649.5979967
   &   464.7577901 &   282.3108313 &   106.5202751 \\
14 &   783.5923916 &   649.4958309 &   509.5689275 &   366.6985461 &   224.2379446
   &    86.2759999 &  &  \\
15 &   355.6845328 &   253.7248107 &   150.7477879 &   50.6624495 &  &  &  &  \\
16 &    75.9322322 &    12.4423751 &  &  &  &  &  &  \\
\hline
 $v$ & $N=16$~~~~~ & $N=17$~~~~~ & $N=18$~~~~~ & $N=19$~~~~~ & $N=20$~~~~~ & $N=21$~~~~~ & $N=22$~~~~~ & $N=23$~~~~~ \\
\hline
 0 & 14624.2491697 & 13913.1247397 & 13185.9781965 & 12445.5582748 & 11694.4913044
   & 10935.2781377 & 10170.2939733 &  9401.7906549 \\
 1 & 12800.3897542 & 12129.4839859 & 11443.8478222 & 10746.1320587 & 10038.8763099
   &  9324.5069207 &  8605.3377477 &  7883.5734587 \\
 2 & 11092.7892574 & 10461.1717306 &  9816.1587298 &  9160.3179022 &  8496.1177758
   &  7825.9269272 &  7152.0161009 &  6476.5630365 \\
 3 &  9499.2238991 &  8906.2090736 &  8301.1932796 &  7686.6775270 &  7065.0772485
   &  6438.7231865 &  5809.8655017 &  5180.6810405 \\
 4 &  8018.1500482 &  7463.3110206 &  6897.9459319 &  6324.5085460 &  5745.3830712
   &  5162.8875786 &  4579.2812539 &  3996.7757693 \\
 5 &  6648.7351228 &  6131.9267980 &  5606.1750566 &  5073.9105382 &  4537.5144973
   &  3999.3262303 &  3461.6556412 &  2926.8019749 \\
 6 &  5390.9100979 &  4912.3055861 &  4426.4853531 &  3935.8889314 &  3442.9337776
   &  2950.0295805 &  2459.6005570 &  1974.1185854 \\
 7 &  4245.4516697 &  3805.5985643 &  3360.4559204 &  2912.5179286 &  2464.2970098
   &  2018.3511303 &  1577.3259174 &  1144.0196903 \\
 8 &  3214.1083701 &  2814.0172644 &  2410.8426901 &  2007.2049589 &  1605.8088269
   &  1209.4988050 &   821.3484334 &   444.8113338 \\
 9 &  2299.7978484 &  1941.0881396 &  1581.9157535 &  1225.1522252 &   873.8793194
   &   531.5178025 &   202.0621779 &  \\
10 &  1506.9327566 &  1192.0932065 &   880.0788887 &   574.2712066 &   278.5621974
   &  &  &  \\
11 &   842.0158921 &   574.9387627 &   315.2186225 &    67.4629279 &  &  &  &  \\
12 &   314.9390144 &   102.3300760 &  &  &  &  &  &  \\
\hline\hline
\end{tabular}
\end{center}
\end{table}

\addtocounter{table}{-1}
\begin{table}[t]
\begin{center}
\caption{(contunued)}
\hspace*{-1mm}\footnotesize
\begin{tabular}{r@{\hspace{4mm}}r@{\hspace{4mm}}r@{\hspace{4mm}}r@{\hspace{4mm}}r@{\hspace{4mm}}r@{\hspace{4mm}}r@{\hspace{4mm}}r@{\hspace{4mm}}r}
\hline\hline
 $v$ & $N=24$~~~~~ & $N=25$~~~~~ & $N=26$~~~~~ & $N=27$~~~~~ & $N=28$~~~~~ & $N=29$~~~~~ & $N=30$~~~~~ & $N=31$~~~~~ \\
\hline
 0 & 8631.9010879 & 7862.6454821 & 7095.9392118 & 6333.6021715 & 5577.3696125
   & 4828.9045738 & 4089.8121877 & 3361.6563825 \\
 1 & 7161.3150780 & 6440.5675977 & 5723.2495883 & 5011.2048856 & 4306.2166211
   & 3610.0241339 & 2924.3437057 & 2250.8947173 \\
 2 & 5801.6598722 & 5129.3231362 & 4461.5065340 & 3800.1170152 & 3147.0350209
   & 2504.1404706 & 1873.3471768 & 1256.6504223 \\
 3 & 4553.2838880 & 3929.7395969 & 3312.0838848 & 2702.3472222 & 2102.5877889
   & 1514.9371687 &  941.6667874 &  385.2907256 \\
 4 & 3417.5514294 & 2843.7793226 & 2277.6516703 & 1721.4242458 & 1177.4779154
   &  648.4128456 &  137.2036513 &  \\
 5 & 2397.0805978 & 1874.8611741 & 1362.6233064 &  863.0411488 &  379.1204527 &  &  &  \\
 6 & 1496.1523329 & 1028.4420536 &  574.0194013 &  136.4148694 &  &  &  &  \\
 7 &  721.4873617 &  313.2174558 &  &  &  &  &  &  \\
 8 &   83.9998731 &  &  &  &  &  &  &  \\
\hline
 $v$ & $N=32$~~~~~ & $N=33$~~~~~ & $N=34$~~~~~ & $N=35$~~~~~  \\
\hline
 0 & 2645.9798564 & 1944.3287761 & 1258.2846381 & 589.5075410 \\
 1 & 1591.4339515 &  947.8028374 &  321.9965509 &  \\
 2 &  656.1964476 &   74.3912135 &  &  \\
\hline\hline
\end{tabular}
\end{center}
\end{table}

\section{Results}

Main results of present work are summarized in two tables. In Table \ref{H2+_res} the resonant states located above the dissociation threshold are given. Generally they are written in the form $E_r+i\Gamma/2$, where $E_r$ is an energy position of the level above the threshold, while $\Gamma$ determines a width of the state. Uncertainty is indicated for the resonance energy only, since the uncertainty for the real and imaginary part is the same in the CCR calculations. If the uncertainty is not shown that implies that all the digits presented are significant. If the imaginary part is omitted then the width of the state is below the uncertainty limit determined by the digits presented in the real part. The most computationally complicate are the states $(v\!=\!18,N\!=\!4)$ and $(v\!=\!17,N\!=\!7)$, where the variational basis of $N_{\rm max}=20\,000$ and $N_{\rm max}=16\,000$ functions have been used. In other cases more moderate basis set of $N_{\rm max}=3000$ to 9000 functions are sufficient.

The bound states of the H$_2^+$ molecular ion are collected in Table \ref{H2+}. All the digits given for a binding energy of a particular state are significant. In fact precision obtained in the numerical calculations is somewhat higher and the numbers shown are taken by truncation of the numerical result to a fixed length.

\subsection{Relativistic and radiative corrections}

Consideration of the relativistic and radiative corrections is essential for comparison with experimental data. Still we intentionally do not present in our work extended sets of numerical results for these corrections, as it was done, say, in \cite{Moss93}. There are several reasons for that.

Generally, for precision spectroscopy aimed for determination of the fundamental constants or for precision tests of the quantum electrodynamics, the states with low $v$ and low $N$ are required. In this case the leading order relativistic \cite{BPrelcor,YanH2+} and radiative \cite{KorobovBL12} corrections are calculated and tabulated for a wide range of vibrational and rotational states. For higher order contributions of orders $m\alpha^6$, $m\alpha^7$, etc, the adiabatic Born-Oppenheimer (BO) approximation may be used \cite{KorobovPRL17}. To this end, results are presented as "effective" potentials \cite{KorobovJPB07,Korobov13}, which then utilized for calculating of corrections as shown in \cite{KorobovPRL17}. Eventually, the theoretical frequencies for particular transitions are compared with precision spectroscopic measurements \cite{Koelemeij07,Koelemeij16,Shen12,Bressel12}. So far such experiments have been performed with HD$^+$ molecular ion only, but even more precise experiments in H$_2^+$ are coming \cite{Shiller17}. In all these cases the hyperfine structure of the states and of experimentally observed spectral lines is of much importance \cite{KorobovPRL06,KorobovPRL16}.

For the case of high $v$ and/or high $N$ states precision of order $10^{-11}$ is not required and adiabatic approximation may be used already for the leading order relativistic correction, which is determined by the Breit-Pauli (BP) Hamiltonian for a bound electron:
\begin{equation}
H_{\rm BP} = -\frac{p^4}{8m^3}
   + \frac{\Delta V}{8m^2} + H_{\rm BP}^{so} + H_{\rm BP}^{ss},
\end{equation}
where $H_{\rm BP}^{so}$ and $H_{\rm BP}^{ss}$ are the electron spin-orbit and electron spin-proton spin interactions, respectively. The "effective" BO potential $\mathcal{E}_{BP}(R)=\alpha^2\left\langle H_{BP} \right\rangle$ may be found in \cite{Howells90,Tsogo06}. Next step is evaluation of the radiative correction at order $m\alpha^5$, this can be done by using the "effective" potential of the leading order radiative correction for a bound electron of the form
\begin{equation}
\mathcal{E}_{\rm SE}^{(5)}(R) = \alpha^3\frac{4}{3}
   \left[\ln{\frac{1}{\alpha^2}}-\beta(R)+\frac{5}{6}-\frac{3}{8}\right]
   \Bigl\langle
      \delta(\mathbf{r}_1)\!+\!\delta(\mathbf{r}_2)
   \Bigr\rangle_{R}\>.
\end{equation}
Here $\beta(R)$ is the Bethe logarithm of a bound electron, and its tabulated data for the two-center problem for a case which corresponds to H$_2^+$ ion may be found in \cite{Korobov13,Kolos92}. For convenience we add to this paper Supplementary Materials \cite{Sup}, which contain the data for the Breit-Pauli relativistic corrections and the nonrelativistic Bethe logarithm for a bound electron in the two-center problem. Beyond that for convenience we have included as well into the Supplementary Materials the Born-Oppenheimer electron energy potential, $\mathcal{E}_{\rm el}(R)$, and the adiabatic corrections, $\mathcal{E}_{\rm ad}(R)$.

\subsection{Conclusion}

In summary, we have computed nonrelativistic energies for all 481 ro-vibrational bound and quasi-bound states in the H$_2^+$ molecular ion, which are supported by the adiabatic $1s\sigma$ potential curve. The calculations are the first \emph{ab initio} non-adiabatic variational calculations, which allowed to get most accurate and complete data for precision spectroscopic studies of the hydrogen molecular ion (cation). We also provide necessary supplemental resources, which may be used to evaluate the relativistic and radiative corrections for individual states as well as for transitions.

\section*{Acknowledgements}

The work has been carried out under financial support of the Russian Foundation for Basic Research under Grant No.~15-02-01906-a.


\begin{thebibliography}{99}

\bibitem{BO} M.~Born and J.R. Oppenheimer, Ann.\ Physik \textbf{84},457 (1927).

\bibitem{BOmc} M.~Born and K.~Huang, \emph{Dynamical Theory oj Crystal Lattices} (Oxford University Press, London, 1954), Appendix VIlI.

\bibitem{Wolnewicz} L.~Wolnewicz and J.D.~Poll, Mol.\ Phys.\ \textbf{59}, 953 (1986); \textbf{66}, 701(E) (1989).

\bibitem{Carrington89} A.~Carrington, I.R. McNab, and Ch.A.~Montgomerie, J.~Phys.~B:\ At.\ Mol.\ Opt.\ Phys.\ \textbf{22}, 3551 (1989).

\bibitem{Moss93} R.E.~Moss, Mol.\ Phys.\ \textbf{80}, 1541 (1993).

\bibitem{Yan99} J. M.~Taylor, Zong-Chao Yan, A.~Dalgarno, and J.F.~Babb, Mol. Phys. \textbf{97}, 25 (1999).

\bibitem{Frolov1995} A.M.~Frolov, and V.H.~Smith, Jr., J.~Phys.~B \textbf{28}, L449 (1995).

\bibitem{var99} V.I.~Korobov, D.~Bakalov, and H.J.~Monkhorst, Phys.\ Rev.~A, {\bf 59}, R919 (1999).

\bibitem{var00} V.I.~Korobov, Phys.\ Rev.~A, {\bf 61}, 064503 (2000).

\bibitem{Hilico00} L.~Hilico, N.~Billy, B.~Gr\'emaud, and D.~Delande, Eur.\ Phys.\ J.~D \textbf{12}, 449 (2000).

\bibitem{Frolov02} D.H.~Bailey and A.M.~Frolov, J.~Phys.~B: At.\ Mol.\ Opt.\ Phys.\ \textbf{35}, 4287 (2002).

\bibitem{Yan07} Hua Li, Jun Wu, Bing-Lu Zhou, Jiong-Ming Zhu, and Zong-Chao Yan, Phys.\ Rev.~A \textbf{75}, 012504 (2007).

\bibitem{Yan14} Ye Ning and Zong-Chao Yan, Phys.\ Rev.~A \textbf{90}, 032516 (2014).

\bibitem{Merkt16} M.~Beyer and F.~Merkt, Phys.\ Rev.\ Lett.\ \textbf{116}, 093001 (2016).

\bibitem{Merkt16JMS} M.~Beyer and F.~Merkt, J.~Mol.\ Spectrosc.\ \textbf{330}, 147 (2016).

\bibitem{CODATA14} P.J.~Mohr, B.N.~Taylor, and D.B.~Newell, Rev.\ Mod.\ Phys.\ \textbf{88}, 035009 (2016).

\bibitem{Var88} D.A.~Varshalovich, A.N.~Moskalev, and V.K.~Khersonskii, \textit{Quantum Theory of Angular Momentum} (World Scientific, Singapore, 1988).

\bibitem{Reinhardt82} W.P.~Reinhardt, Ann.\ Rev.\ Phys.\ Chem.\ textbf{33}, 223 (1982).

\bibitem{Ho83} Y.K.~Ho, Phys.\ Rep.\ {\bf99}, 1 (1983).

\bibitem{ABC} J.~Aguilar and J.M.~Combes, Commun. Math. Phys. \textbf{22}, 269 (1971); E.~Balslev and J.M.~Combes, \emph{ibid.} \textbf{22}, 280 (1971); B.~Simon, \emph{ibid.} \textbf{27}, 1 (1972).

\bibitem{BPrelcor} V.I.~Korobov, Phys.\ Rev.~A \textbf{74}, 052506 (2006).

\bibitem{YanH2+} Zhen-Xiang Zhong, Zong-Chao Yan, and Ting-Yun Shi, Phys.\ Rev.~A \textbf{79}, 064502 (2009).

\bibitem{KorobovBL12} V.I.~Korobov and Zhen-Xiang Zhong, Phys.\ Rev.~A \textbf{86}, 044501 (2012).

\bibitem{KorobovPRL17} V.I.~Korobov, L.~Hilico, and J.-Ph.~Karr, Phys.\ Rev.\ Lett.\ \textbf{118}, 233001 (2017).

\bibitem{KorobovJPB07} V.I.~Korobov and Ts.~Tsogbayar, J.~Phys.~B:\ At.\ Mol.\ Opt.\ Phys.\ \textbf{40}, 2661 (2007).

\bibitem{Korobov13} V.I.~Korobov, L.~Hilico, and J.-Ph.~Karr, Phys.\ Rev.~A, \textbf{87}, 062506 (2013).

\bibitem{Koelemeij07} J.C.J.~Koelemeij, B.~Roth, A.~Wicht, I.~Ernsting, and S.~Schiller, Phys.\ Rev.\ Lett.\ \textbf{98}, 173002 (2007).

\bibitem{Koelemeij16} J.~Biesheuvel, J.-Ph.~Karr, L.~Hilico, K.S.E.~Eikema, W.~Ubachs, and J.C.J.~Koelemeij, Nature Comm.\ \textbf{7}, 10385 (2016).

\bibitem{Shen12} J.~Shen, A.~Borodin, M.~Hansen, and S.~Schiller, Phys.\ Rev.~A \textbf{85}, 032519 (2012).

\bibitem{Bressel12} U.~Bressel, A.~Borodin, J.~Shen, M.~Hansen, I.~Ernsting, and S.~Schiller, Phys.\ Rev.\ Lett.\ \textbf{108}, 183003 (2012).

\bibitem{Shiller17} S.~Schiller, I.~Kortunov, M.~Hern\'andez Vera, F.~Gianturco, and H.~da Silva, Jr.\ Phys.\ Rev.~A \textbf{95}, 043411 (2017).

\bibitem{KorobovPRL06} D.~Bakalov, V.I.~Korobov, and S.~Schiller, Phys.\ Rev.\ Lett.\ \textbf{97}, 243001 (2006).

\bibitem{KorobovPRL16} V.I.~Korobov, J.C.J.~Koelemeij, L.~Hilico, and J.-Ph.~Karr, Phys.\ Rev.\ Lett.\ \textbf{116}, 053003 (2016).

\bibitem{Howells90} M.H.~Howells and R.A.~Kennedy, J.~Chem.\ Soc., Faraday Trans.\ \textbf{86},
3495 (1990).

\bibitem{Tsogo06} Ts.~Tsogbayar and V.I.~Korobov, J.~Chem.\ Phys.\ \textbf{125}, 024308 (2006).

\bibitem{Kolos92} R.~Bukowski, B.~Jeziorski, R.~Moszy\'nski, and W. Ko\l os, Int.~J.~Quantum Chem. \textbf{42}, 287 (1992).

\bibitem{Sup} See Supplemental materials for a set of "effective" potentials for the Breit-Pauli corrections and the nonrelativistic Bethe logarithm.
\end{thebibliography}
\end{document}